# Deformable Molecular Crystal on 2D Crystal: A New Way to Build Nanoscale Periodic Trapping Sites for Interlayer Excitons

Kushal Rijal,[1] Stephanie Amos,[1] Pavel Valencia-Acuna,[1] Fatimah Rudayni,[1,2] Neno Fuller,[1] Hui Zhao[1,*], Hartwin Peelaers[1,*], and Wai-Lun Chan[1,*]

1. Department of Physics and Astronomy, University of Kansas, Lawrence, KS 66045, US

2. Department of Physics, Jazan University, Jazan 45142, Saudi Arabia

*E-mail: huizhao@ku.edu, peelaers@ku.edu, wlchan@ku.edu

**Abstract**

The nanoscale moiré pattern formed at 2D transition metal dichalcogenide crystal (TMDC) heterostructures provides periodic trapping sites for excitons, which is essential for realizing various exotic phases such as artificial exciton lattices, Bose-Einstein condensates, and exciton insulators. At organic molecule/TMDC heterostructures, similar periodic potentials can be formed *via* other degrees of freedom. We utilize the structure deformability of a 2D molecular crystal as a degree of freedom to create a periodic nanoscale potential that can trap interlayer excitons (IXs). Specifically, two semiconducting molecules, PTCDI and PTCDA, which possess similar bandgaps and ionization potentials but form different lattice structures on $MoS_2$, are investigated. The PTCDI lattice on $MoS_2$ is distorted geometrically, which lifts the degeneracy of the two molecules within the crystal's unit cell. The degeneracy lifting results in a spatial variation of the molecular orbital energy, with an amplitude and periodicity of ~ 0.2 eV and ~ 2 nm, respectively. On the other hand, no such energy variation is observed in $PTCDA/MoS_2$, where the PTCDA lattice is much less distorted. The periodic variation in molecular orbital energies provides effective trapping sites for IXs. For IXs formed at $PTCDI/MoS_2$, rapid spatial localization of the electron in the organic layer towards the interface is observed, which demonstrate the effectiveness of these interfacial IX's traps.

Two-dimensional (2D) layered crystals, such as transition metal dichalcogenide crystals (TMDCs), have received much attention recently because different crystals can be stacked together without the constraint imposed by lattice matching between layers, which can result in new material properties.[1] For instance, when two 2D crystals having different azimuthal orientations are stacked together, a moiré pattern can be formed. This moiré pattern typically has a periodicity of a few nm, which can lead to interesting many-body phenomena such as fractal quantum Hall effect[2-4] and unconventional superconductivity in bilayer graphene.[5] In type-II TMDC heterostructures, an electron and a hole are spatially separated by the interface, which can be bound together to become an exciton known as interlayer exciton (IXs) or charge transfer (CT) exciton.[6-8] IXs can have a very long lifetime[9] and they can be formed almost immediately after photoexcitation.[7,10] Hence, IXs play a central role in determining the opto-electronic properties of the heterostructure. Interestingly, a number of recent works have found that the moiré potential present at TMDC/TMDC interfaces can have a strong influence in the emissive properties of IXs,[11-15] as well as their dynamics.[16,17] Moreover, moiré-trapped IXs can be used to realize exotic phases of excitons such as high temperature Bose-Einstein condensates[18] and excitonic insulators,[19] and to provide a platform for quantum information processing.[20] While the moiré potential in TMDC/TMDC heterostructures has attracted much attention, similar nanoscale periodic potentials can also be found at other mixed dimensional vdW heterostructures such as organic/2D heterostructures,[21-23] but have been largely unexplored.

Effective IX formation at organic molecule/TMDC heterostructures have been widely reported in the literature,[24-31] including our previous works.[32-35] However, IX properties at these interfaces are often only investigated as a function of the interfacial band alignment and band bending. The variation in the electronic potential in directions parallel to the interface (the in-plane direction) is often neglected. Organic molecules usually bond to TMDCs *via* weak van der Waals (vdW) interactions. Moreover, molecules can grow epitaxially (known as vdW epitaxy) on an inorganic crystal to form a molecular crystal that has a fixed orientation with respect to the underlying crystal.[36,37] Hence, a periodic nanoscale potential similar to the moiré potential can naturally occur at a molecule/TMDC interface. Here we show that the lattice distortion of the molecular crystal induced by the substrate can induce such nanoscale periodic potential. Unlike TMDCs, a 2D molecular layer is rather "flexible" along any in-plane direction because of the weak vdW bonding between molecules. Hence, a 2D molecular crystal can be distorted easily by the underlying substrate. For example, perylenetetracarboxylic diimide (PTCDI) grown on $MoS_2$ has a distorted in-plane lattice structure with the angle between the two in-plane lattice vectors changed from 90° in the bulk[38] to ~ 83° on $MoS_2$.[39] We show that this lattice distortion, together with the different azimuthal orientations of the two molecules within the crystal's unit cell, can lift the energy degeneracy of molecular orbitals for the two molecules, which produce a spatial variation of the highest molecular occupied orbital (HOMO) and lowest molecular unoccupied orbital (LUMO) energies. By contrast, perylenetetracarboxylic dianhydride (PTCDA) grown on $MoS_2$ has a much less distorted lattice structure. In fact, for PTCDA/$MoS_2$, we do not observe any degeneracy lifting of the HOMO and LUMO levels.

The HOMO and LUMO degeneracy lifting has a significant impact on the IX dynamics and properties. By using a combination of time-resolved optical and photoelectron spectroscopy techniques, we found that in the PTCDI/$MoS_2$ heterostructure, the electron in the IX localizes at the interface due to the strong electron trapping potential at the distorted molecular layer. The electron localization decreases the electron-hole separation, which enhances the photoluminescence (PL) yield of the IX by an order of magnitude. For PTCDI/$MoS_2$, the trapping depth is large (~0.2 eV) and the periodicity is small (~2 nm), which produces a confinement (0.1 eV/nm) that is an order of magnitude stronger than that of the moiré potential typically found at TMDC/TMDC heterostructures (~0.01 eV/nm).[20,40] Hence, such trapping potential would be useful for forming a high density of trapped IXs for realizing correlated excitonic phase such as Bose-Einstein condensates.[18,41] On the other hand, in the PTCDA/MoS$_2$ heterostructure, the electron in the IX is found to delocalize across the PTCDA film. The spatially delocalization increases the average electron-hole distance, which can favor the charge generation from the IX. Such interfaces can be beneficial to applications such as photovoltaics, photocatalysis, and photodetection, where excitons need to be converted into free carriers.

**Energy-level splitting in molecular layer deposited on MoS$_2$**

The electronic structure of the organic layer depends on its lattice structure. Previous scanning tunneling microscopy (STM) work[39] has shown that PTCDI molecules on MoS$_2$ adopt a brick-wall (BW) structure, while PTCDA molecules on MoS$_2$ adopt a herringbone (HB) structure. These two structures are shown in Fig. 1a. The orientation of the lattice vectors for the organic layer (*a$_1$*, *a$_2$*) and the MoS$_2$ (*A$_1$*, *A$_2$*) are also shown in the figure. The structure of our PTCDI/PTCDA layers deposited on MoS$_2$ is measured using low energy electron diffraction (LEED). In these measurements, a ~ 0.7 nm thick film (~ 2 monolayers) was deposited on a bulk MoS$_2$ single crystal. The diffraction patterns for PTCDA and PTCDI are shown in Fig. 1b. The two different patterns show that the two molecules adopt two different structures. Because of the three-fold symmetry of the MoS$_2$ crystal and an azimuthal offset ($\Delta\phi$, defined as the angle between *a$_1$* and *A$_1$*) between the organic and MoS$_2$ crystals, each diffraction pattern is a superposition of diffraction spots from six equivalent molecular domains with different azimuthal orientations. Using these diffraction patterns, we can calculate the lattice constant of the organic layer.

Moreover, the diffraction pattern from the underlying MoS$_2$ layer can be captured from the same sample by using a higher electron energy (Fig. S1 in supplementary information). Comparing the diffraction patterns from the organic layer and the MoS$_2$ allows us to determine the $\Delta\phi$ between the two crystals. The details of the calculation are given in the supplemental materials. The lattice parameters we found are in excellent agreement with the earlier results from the STM measurements (the discrepancy is all within our measurement uncertainties),[39] which are summarized in Table S1 in the supplementary information.

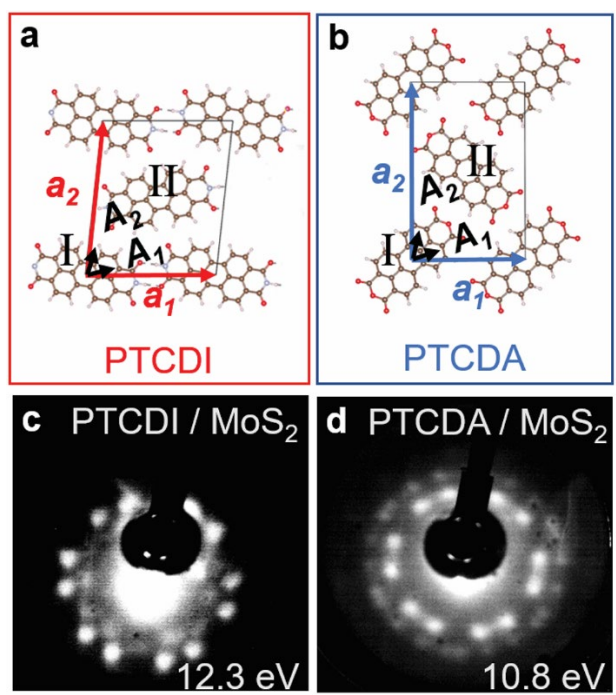

**Fig. 1: 2D lattice structure of molecular layer on MoS$_2$.** (a, b) Structures of PTCDI and PTCDA crystals grown on MoS$_2$. The lattice vectors of the molecular crystals ($a_1$, $a_2$) and that of the underlying MoS$_2$ ($A_1$, $A_2$) are shown. The diagram is drawn based on the lattice structure reported in Ref. [39]. (c, d) LEED images of PTCDI/MoS$_2$ and PTCDA/MoS$_2$. The electron energies used are shown in the figure.

The electronic structures of the two heterostructures were studied by ultraviolet photoemission spectroscopy (UPS) and two photon photoemission spectroscopy (TPPE). In these experiments, molecules are deposited on a CVD-grown monolayer (ML) MoS$_2$ and UPS is used to determine the position of the molecule's HOMO with respect to the Fermi level ($E_f$). A series of UPS spectra for samples with different organic layer thicknesses is shown in Fig. 2a for PTCDI and Fig. 2b for PTCDA. For those sub-nm thick samples, the HOMO position can be heavily influenced by the underlying $MoS_2$ layer. Interestingly, the HOMO peak of the 0.4 nm PTCDI/$MoS_2$ is split into two peaks with a separation of ~ 0.47 eV (the fitting is shown in Fig. S2 in the supplementary information). The two peaks merge back into one as the thickness of the film increases. For thicker samples, the UPS cannot probe molecules near the interface because it is a surface sensitive technique. At the same time, the crystal likely relaxes to its bulk structure as the film becomes thicker. The result indicates that the HOMO peak splitting is an interfacial phenomenon which would be related to the interaction between the PTCDI with $MoS_2$ and/or the BW structure (Fig. 1a) adopted near the interface. On the other hand, no HOMO peak splitting is observed for the 0.5 nm PTCDA/ML $MoS_2$ sample.

The energy offset between the $MoS_2$'s valence-band maximum (VBM) and the molecule's HOMO measured by the UPS, together with the quasiparticle bandgaps reported in the literature,[42-44] are used to determine the band alignment at the interface. The detailed procedures are discussed in the supplementary information, Sec. III. Both interfaces have a similar type-II band alignment (Fig. 2c, d). For an IX forming at the interface, the electron and hole reside in the molecule and the ML-$MoS_2$, respectively. We note that while most previous works have also recognized PTCDA/$MoS_2$ as a type-II heterostructure,[34,45,46] a recent study reported that this interface has a type-I band alignment.[44] To resolve this discrepancy, we follow Ref. [44] and use edges of HOMO and VBM peaks to determine the HOMO-VBM offset. We find that the HOMO-VBM offset is 0.51 eV for our samples, which is larger than the 0.27 eV reported in Ref. [44]. Hence, our interface still has a type-II band alignment. The different band alignments could originate from different sample preparation procedures and substrates used for the TMDC growth.

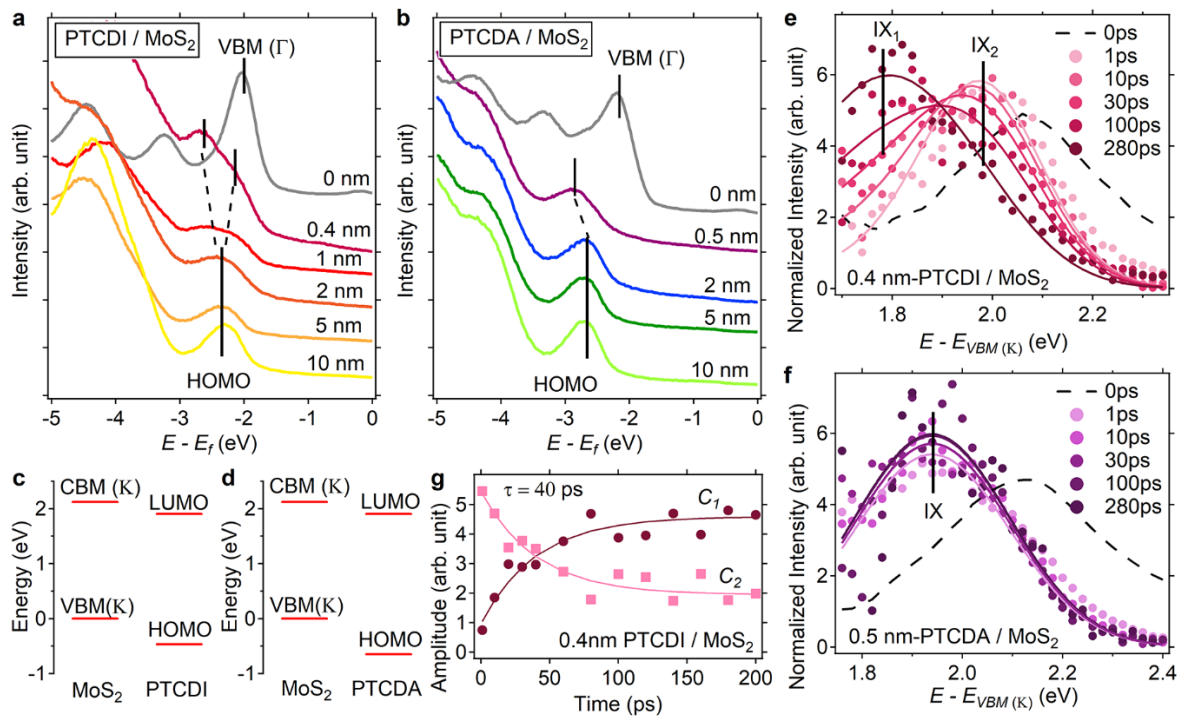

**Fig. 2: Splitting of molecular orbital energies captured by photoemission spectroscopies.** UPS spectra of (a) PTCDI/MoS$_2$ and (b) PTCDA/MoS$_2$ with various organic layer thicknesses. Vertical lines show the molecule's HOMO and MoS$_2$'s VBM at Γ point. HOMO splitting is found in the 0.4-nm PTCDI sample. The energy level diagram (c) PTCDI/MoS$_2$ and (d) PTCDA/MoS$_2$ determined by the measured VBM-HOMO offset, and the quasiparticle bandgaps of the materials. The normalized TPPE spectra for (e) PTCDI/MoS$_2$ and (f) PTCDA/MoS$_2$. For PTCDI/MoS$_2$, two IX peaks are observed and the spectra are fitted with a sum of two Gaussian functions with amplitudes $A_1$ and $A_2$. (g) The amplitude $A_1$ and $A_2$ as a function of time. Both curves are fitted with an exponential rise/decay time of 40 ps (solid lines).

TR-TPPE was used to measure the energies of IXs formed at the interface. In the TR-TPPE experiment, a pump pulse with a photon energy of 1.88 eV was used to selectively excite the MoS$_2$ (PTCDA and PTCDI have an optical band gap of 2.2 eV). Then, a time-delayed probe with a photon energy of 4.68 eV was used to ionize the excited electron at the top organic layer. Our earlier works on PTCDA/ML-MoS$_2$ and PTCDI/ML-MoS$_2$ have shown that, upon the photo excitation of MoS$_2$, charge transfer (CT) from the MoS$_2$ to the molecule occurs in ~ 20 – 30 fs,[34] which produces IXs at the interface. In Figs. 2e and f, the TPPE spectra at selected delay times are plotted as a function of the intermediate state energy referenced with respect to the MoS$_2$'s VBM ($E - E_{VBM(K)}$). The energy scale is chosen because the IX's energy can be estimated by subtracting the intermediate state energy measured by the TPPE with the MoS$_2$'s VBM energy estimated from the UPS measurement. The same method has been previously employed by us[32,33,47-49] and others[50] to estimate the exciton energy from time-resolved photoemission spectra. To compare the spectral shape at different delay times, each spectrum is normalized by its total area underneath the curve. Notably, two peaks can be identified from the PTCDI/MoS$_2$ spectra while only one peak can be found in the PTCDA/MoS$_2$ spectra. The two peaks in the TPPE spectral can be attributed to the splitting of the PTCDI's LUMO.

We note that our method appears to overestimate the IX energy by ~ 0.2 eV. For example, near time-zero (dashed lines in Fig. 2e and f), the TPPE peak locates at ~ 2.1 eV, which is ~ 0.2 eV higher than the pump photon energy (1.88 eV). The error can be attributed to several reasons. First, the MoS$_2$'s VBM is measured before molecule deposition. After the deposition, the VBM position can shift.[51] Second, the TPPE and UPS measurements are done with different light sources. The different excitation conditions can produce different amounts of photo-charging, which can introduce error when the exciton energy is determined from the energy separation of the two peaks located in the two different spectra. If we account for a ~ 0.2 eV overestimation, the IX$_1$ and IX$_2$ peak in PTCDI/ML-MoS$_2$ would have an energy of 1.6 eV and 1.8 eV. This agrees well with our PL measurements, which will be discussed in a later section, where a broad PL peak centered at ~ 1.7 eV is found for the PTCDI/ML-MoS$_2$.

The series of PTCDI/ML-MoS$_2$ spectra shown in Fig. 2e can be fitted with the sum of two Gaussian functions, with the two peaks centered at 1.78 eV and 1.98 eV. It is apparent that the IX relaxes from the higher energy state (IX$_2$) to the lower energy state (IX$_1$) as the time increases. The relaxation time can be determined by plotting the amplitudes of the two Gaussian peaks ($C_1$, $C_2$) as a function of time, which is shown in Fig. 2g. Both curves fit well with an exponential decay/rise function with a time constant of 40 ps. Hence, the data shows a clear population transfer from the IX$_2$ to IX$_1$, and the IX is trapped effectively at the lower energy site. Finally, similar to UPS results, the double peak structure disappears as the organic layer becomes thicker (Fig. S3 in supplemental materials), which shows that the LUMO splitting is also an interface phenomenon, and the traps are located near the interface.

**The physical origin of the energy level splitting**

To determine the physical origin of the HOMO/LUMO splitting, we used density functional theory (DFT) to obtain the electronic structure of a single layer of the PTCDI and the PTCDA crystal. The lattice parameters of the molecular layer reported in Ref. [39] were used as an initial input, which we then refined further by performing full DFT relaxations. Our first-principles calculations of a single PTCDI layer reveal the presence of two peaks in the density of states (DOS) for both HOMO and LUMO levels (Fig. 3a). The wavefunction corresponding to each peak is localized on one of the two molecules of the unit cell as shown in Fig. 3b. For PTCDA, only single peaks for HOMO and LUMO in the DOS occur, but each peak corresponds to two almost degenerate energy levels. This strongly suggests that the double peak structure observed in PTCDI is caused by a splitting of the degenerate energy levels. To distinguish whether the splitting is caused by the molecule itself or by the crystal structure, a calculation is done where PTCDA is placed with the PTCDI crystal structure, and vice versa. PTCDA, which does not show a double peaked DOS in its regular lattice structure, shows a double peaked structure when it is placed with the PTCDI crystal structure. Moreover, when aligning the DOS on the VBM, both HOMO DOS peaks are located at the same energy as those in PTCDI. Similarly, the LUMO also splits. By doing the reverse, that is, placing PTCDI in the PTCDA crystal structure, the double DOS peaks for both HOMO and LUMO disappear. This clearly demonstrates that it is the crystal structure that causes the double peaks, and not the molecule (PTCDI or PTCDA) being used. The HOMO/LUMO splitting found in the calculation is ~ 0.2 eV, which is in reasonable agreement with the experimentally observed splitting (HOMO: 0.47 eV; LUMO: 0.20 eV).

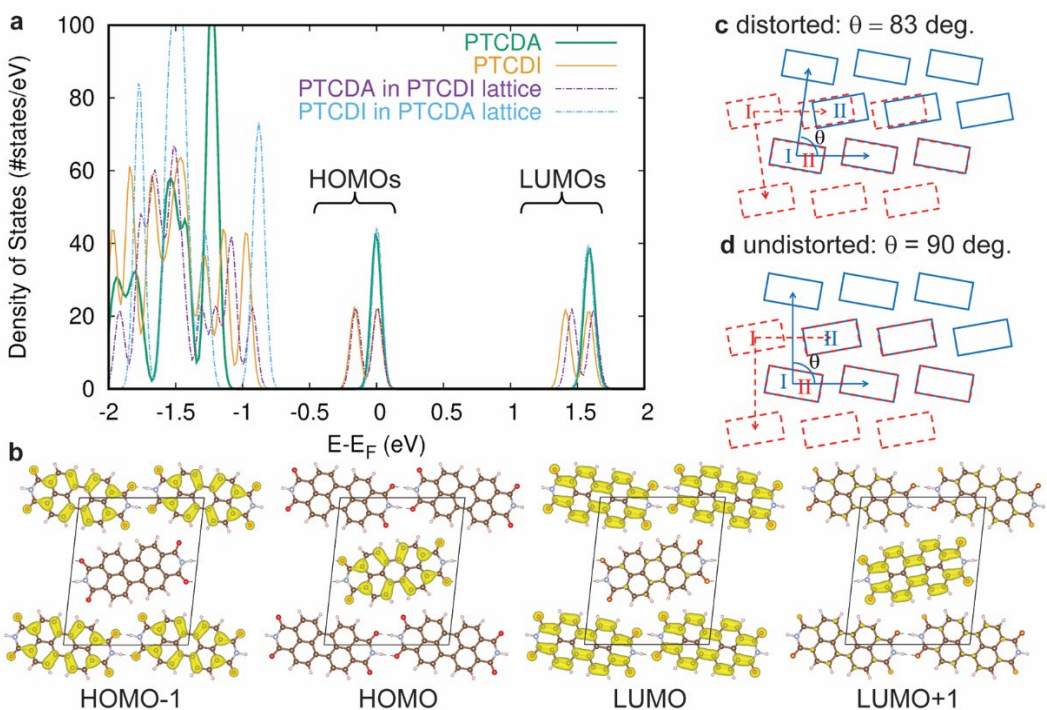

**Fig. 3: Distorted lattice lifts the degeneracy of the HOMO and LUMO of PTCDI.** (a) The DOS of a ML of PTCDA and a ML of PTCDI calculated by DFT. HOMO and LUMO splitting is observed in PTCDI, but not in PTCDA. Placing PTCDI in the PTCDA lattice and vice versa shows that the splitting is caused by the lattice structure. (b) Real space images for the wavefunction of the two HOMOs and two LUMOs of PTCDI shown as isosurfaces corresponding to 10% of the maximum density at $\Gamma$. (c) A schematic showing a distorted lattice makes the two molecules (I and II) in the unit cell having different local environment. (d) An undistorted lattice preserves the mirror symmetry in the lattice, which results in the two molecules (I and II) having the same local environment. In both (c) and (d), the red crystal is a mirror image of the blue crystal by flipping the lattice along the horizontal line.

To further understand why the PTCDI crystal structure, but not the PTCDA crystal structure would lead to HOMO/LUMO splitting, we highlight the similarities and differences between the two structures. Both structures have two molecules in the unit cell, which form a face-centered structure (Fig. 1a). The two molecules within the unit cell (labeled as I and II) have different azimuthal orientations, but they are mirror images of each other. The major difference in PTCDI is that the angle θ between the two lattice vectors ($a_1$ and $a_2$) is ~83°, which is distorted significantly from the 90° in the bulk PTCDI. For comparison, θ is ~89° in PTCDA, which is much closer to 90°. To illustrate how the large lattice distortion in the PTCDI on MoS$_2$ breaks the equivalence of molecules I and II, we flip the crystal vertically. Then, the flipped crystal is translated such that molecule II in the flipped crystal overlaps with molecule I in the original crystal. In Fig. 3c, the original and the flipped PTCDI crystals are shown with blue and red lines, respectively. Molecules I and II have a different local environment as shown by the mismatched locations of blue and red rectangles in the second row. This difference in the local environment can lift the degeneracy of the HOMO/LUMO of molecules I and II. Now, if we keep everything the same except changing θ from 83° to 90° (Fig. 3d), molecules I and II would then have the same environment. We note that keeping θ at 90° allows the lattice to maintain mirror symmetries along both the vertical and horizontal axis. Because molecules I and II are mirror images of each other, maintaining the mirror symmetries of the lattice allows molecules I and II to have equivalent environments. Therefore, we can conclude that the HOMO/LUMO level splitting originates from the distortion of the PTCDI lattice on MoS$_2$, which breaks the mirror symmetry of the lattice and lifts the degeneracy of the two molecules in the unit cell. Finally, we note that θ equals to 90° in bulk PTCDI. We have also calculated the DOS for bulk PTCDI (Fig. S4 in the supplementary information). As expected, we do not observe a double peak structure, but the HOMO and LUMO broaden due to band formation caused by the periodicity.

So far, we have not considered the effect of the $MoS_2$ on the HOMO and LUMO energies. Because the molecular lattice and the $MoS_2$ have different lattice parameters, different molecules will be located at different in-plane locations with respect to the $MoS_2$ unit cell. We estimate the effect of these different environments by determining the energy level alignment at the interface for molecules that are placed in three different locations of the $MoS_2$ unit cell. We find that the variation in the VBM and CBM energies is only ~ 0.05 eV (supplementary information, Sec. VI), which is smaller than the calculated HOMO and LUMO splitting (~ 0.2 eV) in the planar structure. Moreover, variation in the molecular orbital energy due to lattice mismatch produces a continuum of HOMO/LUMO energies, which should result in peak broadening instead of a double-peak structure as observed in our UPS experiment.

**Localization/Trapping of the IX at the Interface**

In practical organic/2D heterostructures, the organic layer is typically thicker than a single layer and the electron within the IX can delocalize across the organic film because of the intermolecular electron coupling. To further understand how the electron delocalization across the organic film would affect the IX trapping, we study the IX's dynamics as a function of the organic film thickness. Because the TR-TPPE technique is only sensitive to electrons residing near the sample surface (for organic materials, the probe depth is ~ 1 – 3 nm[52-54]), the thickness dependent measurement can probe the spatial localization/delocalization of the electron in the IX along the surface normal direction,[47,55] which is parallel to the π-stacking direction of the molecules. Figure 4a and 4b show the temporal evolution of the IX's peak intensity for different film thicknesses for PTCDI/$MoS_2$ and PTCDA/$MoS_2$, respectively. The intensity near time zero is normalized to 1.

For the 1 nm-PTCDI/MoS$_2$ (Fig. 4a, red dots), the intensity decays rapidly in the first few ps and the decay becomes much slower at longer delay times. The initial rapid decay can be attributed to hot exciton/electron relaxation.[49,56,57] The signal at t > 10 ps, on the other hand, should reflect the IX's population. The slow decay, with a decay time constant of 920 ps (dashed line), shows that the IX has a long lifetime, which is a well-known property for IX. Similar dynamics is observed for 1 nm PTCDA/MoS$_2$ (Fig. 2b, dark blue dots), with a decay time constant of 790 ps.

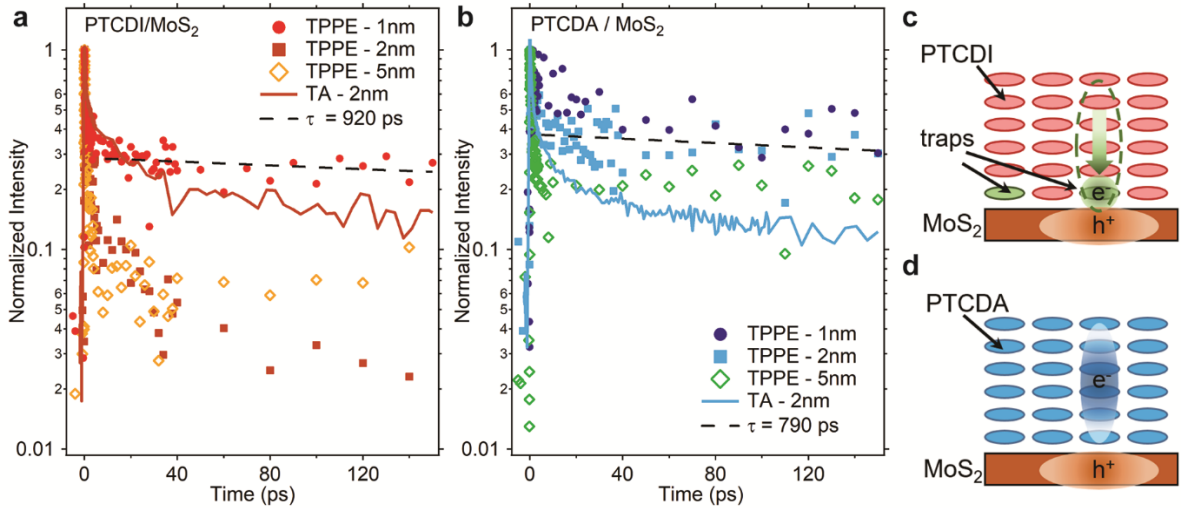

**Fig. 4: Time-resolved measurements reveal the electron localization at the trapping site.** (a) Normalized TPPE (symbols) and TA (solid line) signals for the PTCDI/MoS$_2$ as a function of time. The thickness of the PTCDI film is indicated in the legend. The dashed line indicates a decay time constant of 920 fs. (b) Same as (a), but for PTCDA/MoS$_2$. The dashed line indicates a decay time constant of 790 fs. (c) Electron localization in the PTCDI film leads to the rapid decay in TPPE signal for 2-nm and 5-nm samples. (d) The electron in PTCDA remains delocalized, which results in the similarity of the decay dynamics of the TPPE signal for different PTCDA thicknesses.

Interestingly, the behavior of the two systems deviates when the thickness of the organic film increases. For PTCDI (2 nm, 5 nm)/MoS$_2$, the signal decays much drastically in the first ~50 ps. At first glance, it appears that the IX has a much shorter lifetime in the thicker films as compared to the 1-nm film. However, we note that TPPE is a surface sensitive technique, and it probes the electron density near the surface. Instead of IX recombination, the rapid signal decay can alternatively be attributed to the spatial localization of the electron wavefunction of the IX (see the schematics in Fig. 4c) towards the interface, which decreases the electron density (and, hence, the signal) near the surface with increasing time. To distinguish whether the IX localizes at the interface or recombines, we use transient absorption (TA) measurement to probe the IX's population for the 2-nm sample. In our TA measurement, a pump photon energy of 3.17 eV with a peak fluence of 0.2 µJ cm$^{-2}$ is used, which can excite both layers. A probe photon energy of 1.86 eV is used, which primarily probes the bleaching of the ML-MoS$_2$'s exciton resonance (see supplementary information, Fig. S6). Hence, unlike TPPE which probes the electron in the organic layer, the TA probes the hole in the MoS$_2$. Combining the two techniques gives us a comprehensive view on the temporal evolution of the IX's population.

The normalized TA signals for 2-nm PTCDI/MoS$_2$ and 2-nm PTCDA/MoS$_2$ are shown as solid lines in Fig. 4a and 4b, respectively. For PTCDA/MoS$_2$ (Fig. 4b), the TPPE and TA measurements show similar signal decay rates (the slope of the curve) at $t > 10$ ps. The similarity reaffirms that both measurements are probing the IXs (TPPE probes the electron in PTCDA while the TA probes the holes in MoS$_2$). For PTCDI/MoS$_2$ (Fig. 4a), the TA signal obtained from the 2-nm sample has a slower decay rate as compared to the TR-TPPE signal obtained from the same 2-nm sample. This difference shows that the fast decay rate captured by the TPPE cannot be explained by IX's recombination because recombination should remove the hole in MoS$_2$ as well. On the other hand, the fast decay in the TPPE signal can be explained by the localization of the electron wavefunction towards the interface as mentioned earlier. Under this scenario, the IX still has a long lifetime but its electron is trapped near the interface at longer times. This can result in a fast decay of the TPPE signal because TPPE is a surface sensitive technique and the signal decreases as the electron in the IX is drawn away from the surface. Indeed, for the 1-nm sample, the TPPE can probe electrons residing closer to the interface. In this case, the signal decays much slower, which reflects the real lifetime of the IX.

The spatial localization of the electron can occur when the electron is trapped by a lower energy site near the interface. For PTCDI/MoS$_2$, these lower energy sites (green molecules in Fig. 4c) are produced by the LUMO splitting at the distorted molecular layer near the PTCDI/MoS$_2$ interface. Hence, our results show that the IX can be effectively trapped near the interface even when the organic layer is thick. Practically, a thicker organic layer enables more light absorption, which can favor the production of a high density of IXs near the interface.

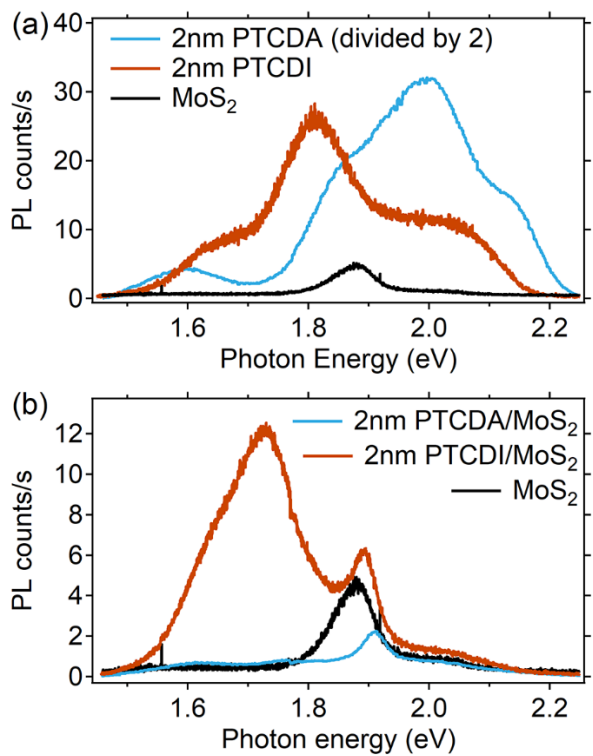

**Fig. 5: Strong IX's PL at the PTCDI/MoS$_2$ caused by IX trapping**. (a) The PL spectra for 2-nm PTCDI, 2-nm PTCDA, and ML-MoS$_2$. (b) The PL spectra for the two heterostructures (PTCDI/MoS$_2$ and PTCDA/MoS$_2$). The strong, red-shifted PL peak in PTCDI/MoS$_2$ is assigned to the PL from the IX.

Under the scenario proposed in Fig. 3d, the IX at the PTCDI/MoS$_2$ interface should have a large radiative recombination rate because the spatial proximity of electron and hole wavefunctions in the IX can enhance the spatial overlapping between the two, which facilitates radiative recombination. To study the radiative recombination of the IX, we measure the steady-state photoluminescence (PL) spectra of the two heterostructures. Samples are excited by a continuous-wave laser beam with a photon energy of 3.05 eV, which is larger than the optical bandgaps of PTCDI, PTCDA, and $MoS_2$. The 1 µW laser beam is focused to a size of about 1 µm in these measurements. For comparison, the PL spectra from individual components that form our heterostructure are also measured under the same conditions, and these spectra are shown in Fig. 5a. All samples used for the PL measurements are fabricated on bulk h-BN. The PL spectra for PTCDA and PTCDI films show a typical three-peak structure. The three-peak structure originates from the hybridization of Frenkel excitons and intermolecular charge transfer excitons in perylene films, together with their vibronic sidebands.[58] The $MoS_2$ shows a single PL peak at 1.88 eV.

The PL spectra for the heterostructures (PTCDI/$MoS_2$, PTCDA/$MoS_2$) are shown in Fig. 5b. Compared with PL spectra from the individual components, a few observations can be made. First, the PL from the organic film is largely quenched (note that the PL counts on the *y*-axis in Fig. 4b is three times smaller than that in Fig. 4a). The PL intensity quenching is especially noticeable for the PTCDA/$MoS_2$. Second, a sharp peak at ~ 1.9 eV on top of a broader PL background is observed. This sharp peak can be attributed to the PL from $MoS_2$. As a comparison, the PL spectrum of the bare $MoS_2$ is also shown in Fig. 5b. Both the height and the width of the residue $MoS_2$'s peak is reduced by ~ half, which indicates that the total PL intensity (the shaded area in Fig. 4b) from $MoS_2$ is reduced in the heterostructure as compared to the bare $MoS_2$ sample. The quenching of the PL signal from both the organic film and the $MoS_2$ can be attributed to the interfacial electron and hole transfer, respectively. We note that both the excitation photon energy (3.06 eV) and the photon energy (1.88 eV) of the $MoS_2$'s PL are outside the absorption window of PTCDA and PTCDI. At 3.06 eV, the optical absorption coefficient for PTCDA and PTCDI is ~ $5 \times 10^4$ cm$^{-1}$. For a 2-nm film, only 1% of the light intensity is absorbed. The absorption coefficient at 1.88 eV for the two molecules is even weaker (< $1 \times 10^4$ cm$^{-1}$). Hence, the quenching of the MoS$_2$'s PL signal cannot be explained by the optical absorption from the organic layer. Third, a rather broad PL peak at 1.72 eV can be found in the PTCDI/MoS$_2$. This PL peak shows a clear red shift as compared to the PL peak of the bare PTCDI sample shown in Fig. 5a. The energy of this PL peak roughly agrees with the energy of the IX$_1$ peak in the TPPE spectra shown in Fig. 2e, considering that the TPPE/UPS method appears to overestimate the exciton energy by ~0.1 – 0.2 eV. Hence, we assign the PL peak at ~ 1.7 eV to the IX exciton. For the PTCDA/MoS$_2$, although a small shoulder is found at ~ 1.6 eV, the PL intensity is at least an order of magnitude weaker as compared to that of PTCDI/MoS$_2$.

Typically, the PL yield is proportional to $k_r/(k_{nr}+k_r)$, where $k_r$ and $k_{nr}$ are radiative and non-radiative recombination rates, respectively. The IX has a low PL yield, which means $k_{nr} \gg k_r$. Hence, the PL intensity is roughly proportional to $k_r/k_{nr}$. However, as long as $k_{nr} \gg k_r$, a larger $k_r$ cannot significantly change the overall IX's lifetime ($1/(k_{nr}+k_r) \sim 1/k_{nr}$). Indeed, the IXs at PTCDI/MoS$_2$ and PTCDA/MoS$_2$ have similar lifetimes (Figs. 4a and 4b). On the other hand, the PL yield of the IX at PTCDI/MoS$_2$ is at least an order of magnitude larger than that of the IX at PTCDA/MoS$_2$. Therefore, the time-resolved and the PL measurements together show that the IX at PTCDI/MoS$_2$ has a much larger $k_r$, which is consistent with our earlier claim that the electron in the IX is trapped near the interface (Fig. 4c).

In conclusion, we show that the deformability of a 2D-vdW bonded molecular layer can be used as a degree of freedom to produce an array of periodic interfacial traps at organic/TMDC interfaces. For the PTCDI/MoS$_2$ heterostructure, the distorted PTCDI lattice breaks the mirror symmetry in the lattice which lifts the degeneracy of the two molecules in the unit cell. The variation in the IX energy between neighboring molecular sites is as large as 0.2 eV, which can confine the IX within a single molecule. Because one in every two molecules in the molecular crystal is an IX trapping site, a maximum trap density of ~ $4\times10^{13}$ cm$^{-2}$ can be achieved. The density is around an order of magnitude larger than that achieved by the moiré pattern at TMDC/TMDC interfaces (~$10^{12}$ cm$^{-2}$).[20,40] A high IX density favors the formation of correlated phases such as Bose-Einstein condensates. Moreover, different molecules can form lattice with different symmetries, which enables the formation of a large varieties of exciton lattices.

## METHODS

### Sample Preparations

Two kinds of monolayer (ML)-MoS$_2$ were used. CVD-grown ML-MoS$_2$ was purchased from a commercial vendor (6 carbon technology). These MoS$_2$ samples were grown on SiO$_2$/Si substrates, and they were used in the UPS, TPPE, and transient absorption experiments. For the LEED measurement, single crystal, bulk-MoS$_2$ (SPI Supplies) was used.

For the PL measurements, exfoliated MoS$_2$ flakes were used. The MoS$_2$ and h-BN flakes were mechanically exfoliated from bulk crystals using adhesive tape and subsequentially transferred onto polydimethylsiloxane (PDMS) substrates. Monolayer flakes were identified by their known contrasts and were transferred onto desired substrates by a micromanipulator under an optical microscope with a long working distance. The monolayer thickness of MoS$_2$ was also confirmed by PL measurements.

Thin films of PTCDA (Alfa Aesar, 98%) and PTCDI (Luminescence Technology, > 99%) with different thicknesses (0.4 – 5 nm) were deposited on MoS$_2$. The MoS$_2$ was annealed in the ultrahigh vacuum (UHV) chamber with a base pressure of $1 \times 10^{-10}$ Torr overnight at 420 °C before the deposition. The cleanliness of the MoS$_2$ surface was verified by the observation of a clear band structure with angle-resolved photoemission spectroscopy (ARPES).[32] After annealing, the sample was transferred *in-situ* to a UHV deposition chamber with a base pressure of 1 × 10$^{-9}$ Torr. The thickness of the organic thin film was monitored by a quartz crystal microbalance. For thicknesses less than or equal to 2 nm, the molecules were deposited at a rate of 0.3 Å/min at a sample temperature of 90 °C. For thicknesses greater than 2 nm, molecules were deposited at a rate of 0.7 Å/min with the sample kept at the room temperature.

**UPS experiment**

A standard UV lamp was used to produce the UV light for the UPS measurement. The He-I emission line having 21.22 eV photon energy was used for our experiment. The energy and angle of emission of the photoelectrons were measured with a hemispherical analyzer equipped with an imaging detector (Phoibos 100, SPECS).

**TR-TPPE experiment**

The TR-TPPE is a pump-probe technique in which the probe pulse is used to obtain the photoemission spectrum. The pump beam had a photon energy of 1.88 eV, while the probe beam had a photon energy of 4.63 eV. A Yb: KGW regenerative amplifier running at 125 kHz (Pharos, 10 W, Light Conversion) was used to pump two noncollinear optical parametric amplifiers (Orpheus-2H, Orpheus-3H, Light Conversion) whose output generated our pump and probe beams. The width of pump and probe pulse were 25 fs and 65 fs, respectively. Pulses at the sample have a Gaussian profile with a full width half maxima (fwhm) size of ∼0.8 mm. The kinetic energy of the emitted photoelectrons was measured by the hemispherical electron analyzer mentioned earlier. All measurements were taken at room temperature. The pulse energy for the pump and probe pulses were 280 nJ and 10 nJ, respectively.

**Photoluminescence spectroscopy**

The PL spectroscopy measurements were performed with a 405-nm continuous-wave diode laser as the excitation source. It was focused with a microscope objective lens to a spot size of about 1.0 μm. An imaging system is used to monitor the sample surface and the excitation laser spot for beam alignment. The PL is collimated by the same objective lens and directed to a spectrometer (HORIBA Scientific, iHR550), equipped with a thermoelectrically cooled charge-coupled device (CCD) camera.

**Transient absorption**

Transient absorption (TA) measurements were performed with a homemade setup. An optical parametric oscillator was pumped by a 80-MHz Ti:sapphire oscillator, producing a signal output around 1320 nm. Its second harmonic, generated in a nonlinear crystal, was used as the probe pulse. The pump pulse, with a wavelength of about 391 nm, was obtained by second-harmonic generation of the Ti:sapphire output. The pump and probe pulses were combined by a beamsplitter and co-focused to the sample by a microscope objective lens, with spot sizes of about 1.8 μm. The reflected probe by the sample was detected by a silicon photodiode and a lock-in amplifier. A mechanical chopper was used in the pump arm to modulate its intensity at about 3 kHz. With this setup, we measured the differential reflectance of the probe, which is defined as $\Delta R/R_0 = (R-R_0)/R_0$, where $R$ and $R_0$ are the probe reflectance with and without the presence of the pump beam, respectively. Under the experimental conditions of this study, the differential reflectance is proportional to the carrier density in $MoS_2$ since the probe photon energy was tuned to its optical bandgap. Hence, the time evolution of the differential reflectance, measured by scanning the time delay of the probe pulse with respect to the pump pulse, reveals the dynamics of carrier density in $MoS_2$.

**DFT calculations**

All DFT calculations were performed using projector augmented wave (PAW) potentials[59] as implemented in the Vienna Ab-initio Simulation Package (VASP).[60,61] We use the SCAN meta-GGA functional[62] and explicitly included van der Waals interactions using the Grimme-D3[63] method. The plane wave expansion cutoff was set to 525 eV, and all structures were relaxed so that the forces were smaller than 20 meV/A and stresses smaller than 0.04 meV/A$^3$. We sampled reciprocal space using a 10×10×1 Γ-centered *k*-point grid.

ACKNOWLEDGEMENTS

This work is primarily supported by US National Science Foundation Grant No. DMR-2109979. W. –L. C. also acknowledges the support from University of Kansas General Research Fund (2151080). H. Z. is supported by the U.S. Department of Energy, Office of Basic Energy Sciences, Division of Materials Sciences and Engineering under Award DE-SC0020995. F.R. acknowledges scholarship support from Jazan University. The calculations were performed at the University of Kansas Center for Research Computing (CRC), including the BigJay Cluster resource funded through NSF Grant MRI-2117449.

AUTHOR CONTRIBUTIONS

K.R., F.R. and N.F., supervised by W. L. C., performed the photoemission spectroscopy, LEED measurements and organic film deposition. P.V.A., supervised by H. Z., performed the transient absorption and photoluminescence measurements. K.R. coordinated all the experimental efforts. S.A., supervised by H. P., performed DFT calculations. W.L.C., with the help from H. P. and H.

Z., conceived the research idea. K. R., S. A., P. V. A., H. Z., H. P., and W. L. C. prepared the manuscript.

Supplementary Information

# Deformable Molecular Crystal on 2D Crystal: A New Way to Build Nanoscale Periodic Trapping Sites for Interlayer Exciton

Kushal Rijal,[1] Stephanie Amos,[1] Pavel Valencia-Acuna,[1] Fatimah Rudayni,[1,2] Neno Fuller,[1] Hui Zhao[1],*, Hartwin Peelaers[1],*, and Wai-Lun Chan[1], *

1. Department of Physics and Astronomy, University of Kansas, Lawrence, KS 66045, US

2. Department of Physics, Jazan University, Jazan 45142, Saudi Arabia

*E-mail: huizhao@ku.edu, peelaers@ku.edu, wlchan@ku.edu

# I. LEED Measurements

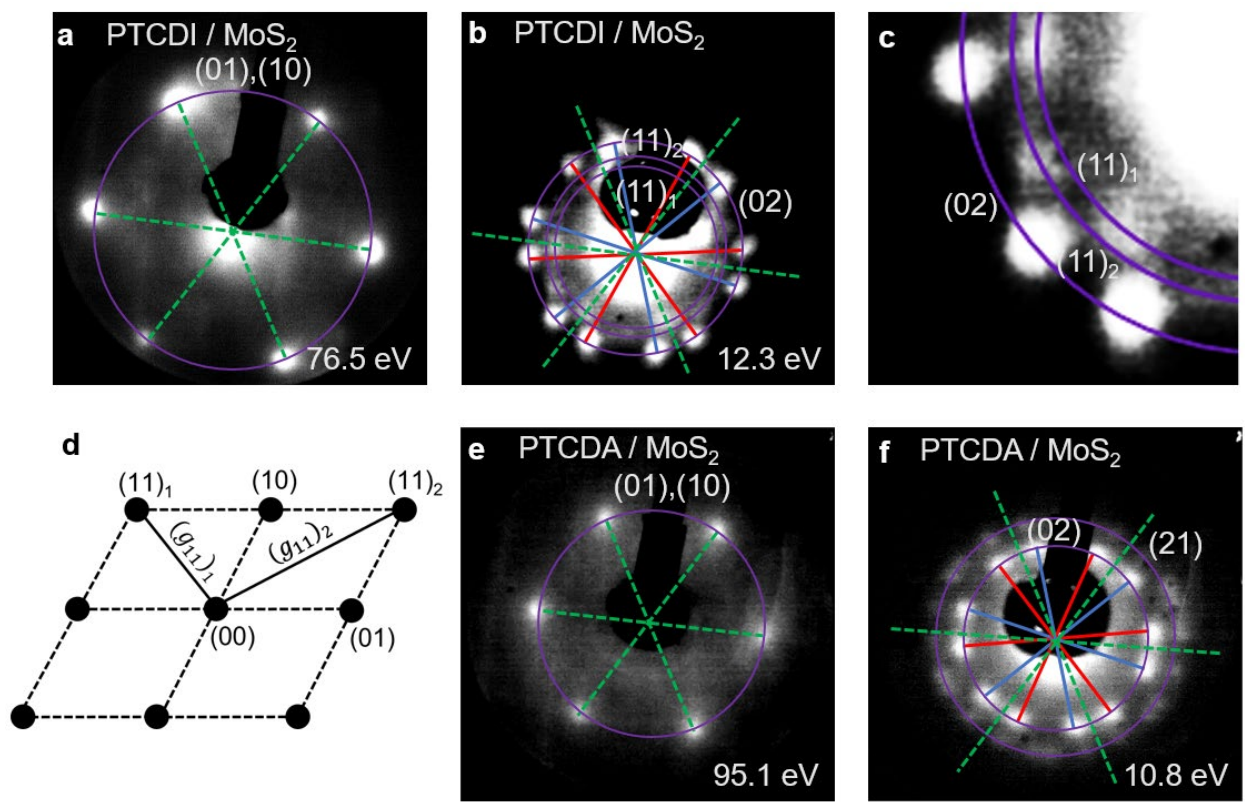

Fig. S1: LEED images of (a) PTCDI/MoS$_2$ at 76.5 eV and (b) PTCDI/MoS$_2$ at 12.3 eV. (c) A section of (b) is enlarged to show that (1 1) diffraction spots can be grouped into two rings with different radii. (d) A schematic diagram illustrates the origin of the two (1 1) rings when the lattice is distorted from a rectangular lattice. LEED images of (e) PTCDA/MoS$_2$ at 95.1 eV and (f) PTCDA/MoS$_2$ at 10.8 eV. In all LEED images the directions of the reciprocal lattice vectors of MoS$_2$ are shown as green dashed lines. The thickness of the PTCDI and PTCDA films used in the measurement are 0.6 nm and 1 nm, respectively.

LEED probes the surface of the sample due to its low penetration depth and it measures the reciprocal lattice in a 2D plane that is parallel to the sample surface. For all LEED measurements, molecules are deposited on a bulk MoS$_2$ single crystal. Diffraction patterns collected at low electron energies (Fig. S1b, Fig. S1f) reveal the PTCDI and PTCDA reciprocal lattices. The large number of diffraction spots (each ring has either 12 or 24 spots) is originated from six structurally equivalent domains with different azimuthal orientations, which is originated from the 3-folded symmetry and a mirror symmetry of the MoS$_2$ lattice. There are six instead of three equivalent domains because the lattice vector of the molecular lattice ($\vec{a_1}$) is not in the same direction as the lattice vector of MoS$_2$ ($\vec{A_1}$). We define the offset angle between $\vec{a_1}$ and $\vec{A_1}$ as Φ, and the definition of all lattice vectors are shown schematically in Fig. 1 in the main text. The LEED image (Fig. S1f) of the PTCDA/MoS$_2$ shows 2 rings: an inner ring with (0 2) diffraction spots and an outer ring with (1 2) diffraction spots. On the other hand, the LEED image (Fig. S1b) of the PTCDI/MoS$_2$ shows two rings of (1 1) spots (with slightly different diameters) and one ring of (0 2) spots. The presence of two diffraction (11) diffraction rings (Fig. S1c) shows that the angle (defined as θ) between the PTCDI lattice vectors ($\vec{a_1}$ and $\vec{a_2}$) deviates from 90°. As shown schematically in Fig. S1d, if the lattice is distorted from a rectangle to a parallelogram, the (1 1) diffraction spots can be grouped into two types of reciprocal lattice vectors, $(\vec{g_{11}})_1$ and $(\vec{g_{11}})_2$, with slightly different lengths. For the diffraction pattern of PTCDA, we can fit all the (1 2) spots into a single ring, which indicates that θ is close to 90°.

The orientation Φ of the molecular lattice with respect to the MoS$_2$ lattice is determined by the offset angle between reciprocal lattice vectors from the PTCDI/PTCDA diffraction pattern obtained at low energies (Fig. S1b, S1f) and the MoS$_2$ diffraction pattern obtained at high energies (Fig. S1a, Fig. S1e). Lattice parameters, $\vec{a_1}$, $\vec{a_2}$ and θ, for the PTCDI molecules were calculated using the measured lengths in the reciprocal space of the three diffraction rings $(g_{02})$, $(g_{11})_1$, and $(g_{11})_2$ and the following three equations:

$$g_{02} = 2 \times g_{01} = \left(\frac{4\pi}{a_2 \sin\theta}\right)$$

$$((g_{11})_1)^2 = g_{01}^2 + g_{10}^2 - 2g_{01}g_{10}\cos\theta = \left(\frac{2\pi}{a_2 \sin\theta}\right)^2 + \left(\frac{2\pi}{a_1 \sin\theta}\right)^2 - \frac{8\pi^2 \cos\theta}{a_1 a_2 \sin^2\theta}$$

$$((g_{11})_2)^2 = g_{01}^2 + g_{10}^2 + 2g_{01}g_{10}\cos\theta = \left(\frac{2\pi}{a_2 \sin\theta}\right)^2 + \left(\frac{2\pi}{a_1 \sin\theta}\right)^2 + \frac{8\pi^2 \cos\theta}{a_1 a_2 \sin^2\theta}$$

Similarly, the lattice parameters, $\vec{a_1}$ and $\vec{a_2}$, of the PTCDA lattice are calculated by using the lengths of the (0 2) and (2 1) diffraction rings in the reciprocal space and the equations below. For the PTCDA lattice, Γ was taken to be 90° as we cannot distinguish the two (2 1) rings with the resolution of our LEED image.

$$g_{02} = 2 \times g_{01} = \left(\frac{4\pi}{a_2 \sin\theta}\right)$$

$$g_{12}^2 = g_{10}^2 + g_{02}^2 = \left(\frac{2\pi}{a_1 \sin\theta}\right)^2 + \left(\frac{4\pi}{a_2 \sin\theta}\right)^2$$

The lattice parameters we obtained are summarized in the table below. We also compared the parameters we obtained with those reported in Ref. [S1].

|  | PTCDI | | PTCDA | |
| --- | --- | --- | --- | --- |
|  | LEED | STM (Ref. S1) | LEED | STM (Ref. S1) |
| $|\vec{a_1}|$ | 14.1 ± 0.6 Å | 14.5 Å | 12.5 ± 0.6 Å | 12.4 Å |
| $|\vec{a_2}|$ | 17.1 ± 0.2 Å | 17.2 Å | 20.0 ± 0.2 Å | 19.7 Å |
| Molecules per unit cell | 2 | 2 | 2 | 2 |
| $\theta = \angle(\vec{a_1}, \vec{a_2})$ | 85.1° ± 2.2° | 83.1° | 90° | 88.8° |
| $\Phi = \angle(\vec{A_1}, \vec{a_2})$ | 11.2° ± 0.7° | 10.9° | 12.1° ± 0.8° | 12.7° |

## II. UPS peak fitting

In order to determine the location of the two HOMO peaks in the 0.4 nm-PTCDI/MoS$_2$ ultraviolet photoemission spectroscopy (UPS) spectrum, we subtract the background originated from inelastic electrons (Fig. S2a). Then, the subtracted spectrum is fitted with two Gaussian peaks (Fig. S2b). It is found that the two HOMO peaks are separated by 0.47 eV.

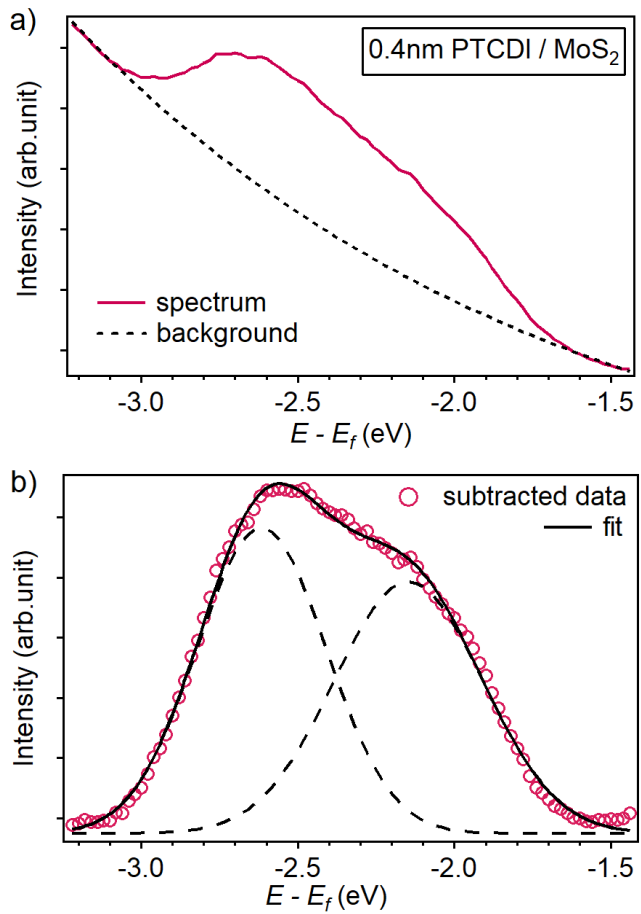

Fig. S2: (a) The raw UPS spectrum (solid line) together with the background from inelastic electrons (dashed line). (b) The background is subtracted from the raw data (open circle). The subtracted data is fitted with a double Gaussian function.

**III. Procedures in determining the energy level alignment**

The energy level alignment at the organic/MoS$_2$ interface is determined by using UPS. For the measurement shown in Fig. 2a and 2b in the main text, we collected the photoelectrons along the surface normal direction, which corresponds to the Γ point in the *k*-space. Unlike bulk crystals, monolayer (ML) TMDC does not have dispersion along the surface-normal direction. Therefore, we simply use the peak position of the VBM peak as the valence-band maxima (VBM) at the Γ point. The actual VBM for ML-MoS$_2$ is at the K point. For ML-MoS$_2$, the VBM at the K point is 0.13 eV higher than that at the Γ point.[S2] This difference is added to the VBM at the Γ point to determine the VBM at the K point.

For determining the positions for PTCDA and PTCDI HOMO, the spectra obtained from 2-nm samples are used. Because the bandwidth of these molecular crystals (< 200 meV)[S3] is much narrower than the width of the photoemission peak (1 eV), most of the peak's width is caused by inhomogeneous broadening. Therefore, we use the peak position as the HOMO position. We note that using the edge-to-edge HOMO-to-VBM offset rather than the peak-to-peak HOMO-to-VBM offset does not significantly change the result. For PTCDA/MoS$_2$, the peak-to-peak and edge-to-edge offsets are 0.52 eV and 0.51 eV, respectively. For PTCDI/MoS$_2$, the peak-to-peak and edge-to-edge offsets are 0.34 eV and 0.2 eV, respectively. In both cases, both interfaces still have a type-II band alignment. In our energy level diagram, we use the peak-to-peak HOMO-to-VBM offset.

The positions of the molecules' HOMO and the MoS$_2$'s VBM at the K point are shown in Fig. 2c and Fig. 2d in the main text. In these energy level diagrams, the HOMO level of the molecules is set at 0 eV. In order to determine the position of the conduction-band minimum (CBM) and the lowest unoccupied molecular orbital (LUMO), reported bandgaps of these materials are used. We note that these materials have a large exciton binding energy. For ML-MoS$_2$, the quasiparticle bandgap can be found by adding the optical bandgap (1.88 eV) to the exciton binding energy of 0.24 eV.[S4] For molecules, the transport gaps estimated from UPS and inverse photoemission spectroscopy (IPES) are used. Transport gaps for PTCDA and PTCDI are 2.55 eV[S5] and 2.37 eV,[S6] respectively.

"Peak-to-peak" versus "Edge-to-edge" in finding exciton energy

Finally, we note that the higher energy edge of the HOMO peak and the lower energy edge of the LUMO peak are often used in UPS-IPES measurements to estimate the transport gap. In the case of UPS-IPES, it is known that the peak-to-peak separation can overestimate the transport gap by as large as 1 eV because of the polarization effect.[S6] In UPS (IPES) experiment, an electron is removed (added) to the surface, which forms a positively-charged (negatively-charged) polarization cloud on the surface. The polarization effect shifts the UPS and IPES spectra in opposite direction, and can overestimate the bandgap by as much as 0.8 eV.[S6] Hence, the edge-to-edge method is often used to artificially "correct" for this overestimation.[S6] In our case, TPPE instead of IPES is used to probe the excited state. In contrary to IPES, electrons *are removed* from the surface in the TPPE experiment. Hence, the polarization effect will shift the TPPE and UPS spectra in the same direction, which should have much less effect on the energy distance between a UPS peak and a TPPE peak. Therefore, in our case, it is more appropriate to use peak-to-peak separation when a pair of UPS and TPPE spectra are used to find the exciton energy.

## IV. Additional TPPE spectra

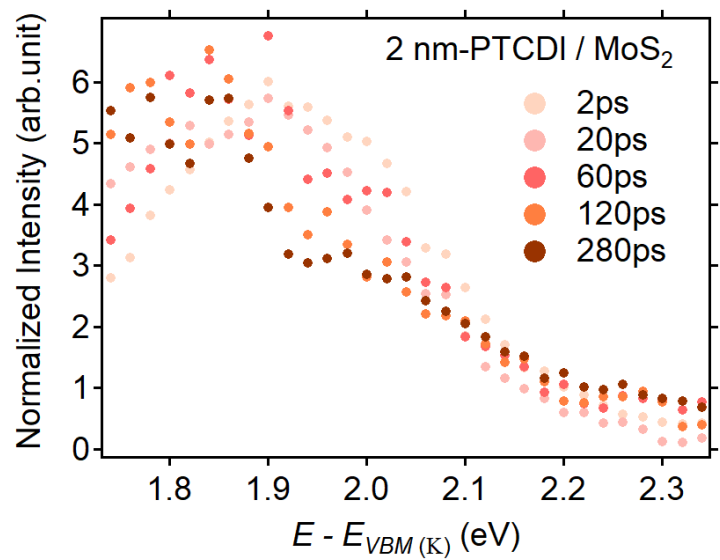

Fig. S3: The normalized TPPE spectra for 2nm-PTCDI/MoS$_2$. Unlike the 0.4 nm sample, only one IX peak is observed.

## V. DFT calculations for bulk PTCDI and PTCDA

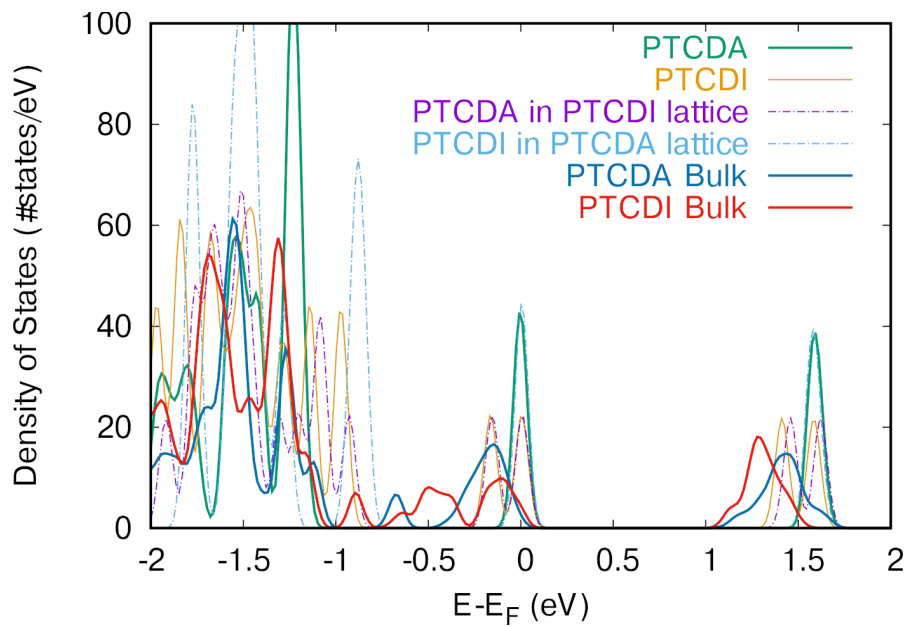

Fig. S4: The DOS of bulk PTCDA and bulk PTCDI calculated by DFT are shown together with the 2D molecular lattice discussed in the main text. For the bulk calculations, we started with the experimental structures as reported in Ref. [S7,8] as initial structures that were fully relaxed using DFT.

## VI. Effects of the MoS₂ on VBM and CBM energies

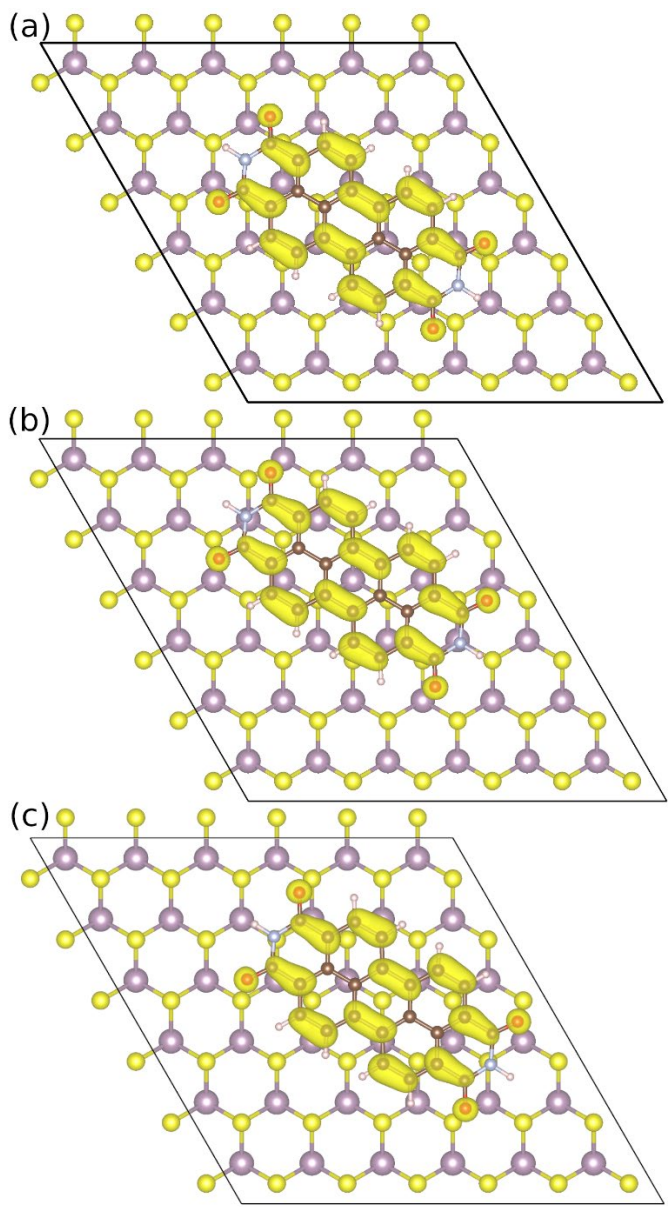

Fig. S5: To estimate the effect of the substrate, the CBM and VBM energies of the heterostructure are calculated by replacing the molecule on top of the (a) empty site, (b) Mo site, and (c) S site of the MoS$_2$ crystal. The calculated VBM, CBM and band gap energies are shown in Table S2. The Grimme-D2[S9] method is used to describe the van der Waals interactions, as this method works best for MoS$_2$,[S10] while the differences for the PTCDI molecule between D2 and D3 are minimal. *K*-space was sampled by the Γ point.

| Site | PTCDI | | PTCDA | |
| --- | --- | --- | --- | --- |
| | VBM (eV) | CBM (eV) | VBM (eV) | CBM (eV) |
| a – empty site | -5.98 | -4.42 | -6.16 | -4.66 |
| b – Mo site | -5.97 | -4.44 | -6.15 | -4.68 |
| c – S site | -6.02 | -4.48 | -6.15 | -4.71 |

Table S2: VBM and CBM energies (in eV) of the heterostructure when the molecule is placed at different locations on the MoS$_2$ crystal (the 3 different locations are shown in Fig. S5). To be able to compare all structures, we aligned the valence and conduction band edges using the vacuum energy as an absolute reference.

## VII. Control experiments for the TA measurement

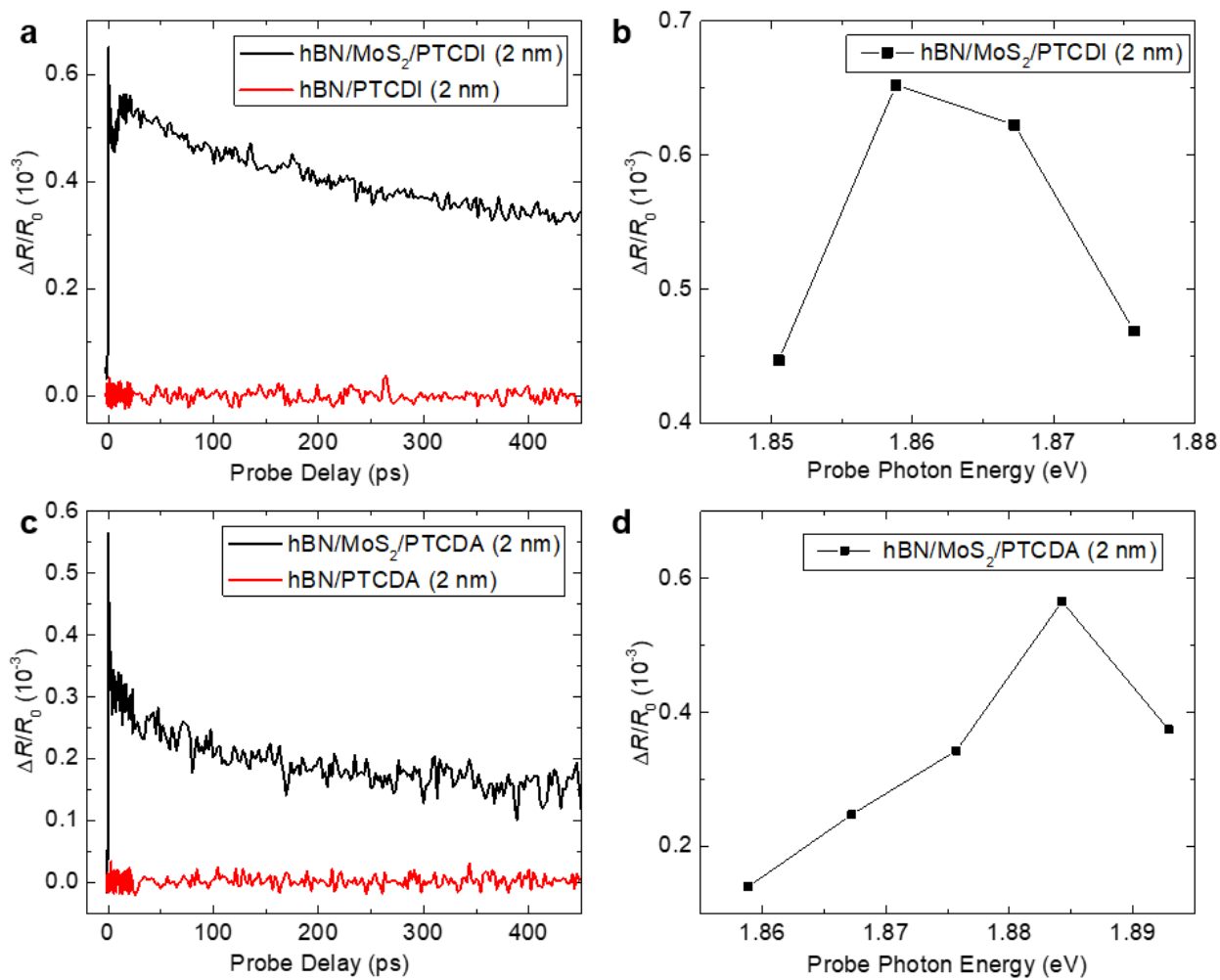

Fig. S6: (a) The transient absorption measurements for 2 nm-PTCDI/MoS$_2$ and 2 nm-PTCDI. The comparison shows that the 1.86-eV probe is only sensitive to excited carriers in MoS$_2$. (b) The signal amplitude as a function of the probe photon energy for the 2 nm-PTCDI/MoS$_2$ sample. The peak corresponds to the excitonic resonance of MoS$_2$. (c), (d) Same as (a) and (b), except PTCDI is replaced by PTCDA.

To further confirm that our transient absorption signal monitors carriers in MoS$_2$ and rule out potential contributions from the molecular layers, two control samples were fabricated and studied. Each sample contains regions of MoS$_2$/molecular layer heterostructure and molecular layer alone, facilitating their direct comparison. To fabricate the first sample, a thick hexagonal boron nitride (hBN) flake was first exfoliated from a bulk crystal and transferred to a Si/SiO$_2$ substrate. A monolayer MoS$_2$ flake is then exfoliated and transferred onto the hBN flake, covering part of it. Then, a 2-nm PTCDI layer is deposited to cover the entire substrate.

We first repeat the transient absorption measurements discussed in the main text on the region of this sample where PTCDI was directly on top of hBN. With the same experimental conditions as used in the main text, Figure 4a, (3.17-eV and 0.2 μJ cm-2 pump and 1.86-eV probe), no signal was observed, as indicated by the red curve in Figure S6(a). Since the pump is capable of exciting carriers in PTCDI, this experiment proves that the transient absorption signal of the 1.86-eV probe cannot originate from carriers in PTCDI. Furthermore, moving the laser spots to the region of hBN/MoS$_2$/PTCDI, a signal with similar magnitude and temporal behavior as main text, Figure 4a, was observed [black curve in Figure S6(a)]. Furthermore, the signal magnitude shows a strong dependence on the probe photon energy near the excitonic resonance of MoS$_2$, as shown in Figure S6(b).

The second control sample is similar to the first one, except that the molecular layer is PTCDA. The control measurements were done in the same way as the first control sample. The results are summarized in Figure S6(c) and S6(d). These results solidify that the transient absorption signal shown in Figure 4(b) of the main text monitors the holes in MoS$_2$.

S12